\documentclass{elsart} 
\usepackage{graphicx} 
\usepackage{amssymb} 
\begin{document} 

\def\eeq{\end{equation}} 
\def\beq{\begin{equation}} 
\def\bea{\begin{eqnarray}} 
\def\eea{\end{eqnarray}} 

\begin{frontmatter}  

\title{Mixing and relaxation dynamics of the Henon map at the edge of chaos}  
\author{Ernesto P. Borges} 
\address{Escola Politecnica, Universidade Federal da Bahia, Rua Aristides  
Novis 2, 40210-630 Salvador BA, Brazil \\} 
\author{Ugur Tirnakli\corauthref{cor}}  
\corauth[cor]{Corresponding author.} \ead{tirnakli@sci.ege.edu.tr} 
\address{Department of Physics, Faculty of Science, Ege University, 35100 Izmir, Turkey}   

\begin{abstract} 
The mixing properties (or sensitivity to initial conditions) and relaxation  
dynamics of the Henon map, together with the connection between these concepts,  
have been explored numerically at the edge of chaos. It is found that the  
results are consistent with those coming from one-dimensional dissipative maps.  
This constitutes the first verification of the scenario in two-dimensional  
cases and obviously reinforces the idea of weak mixing and weak chaos.\\ 
\end{abstract}  

\begin{keyword} Nonextensive thermostatistics, Henon map, dynamical systems 
\end{keyword} 

\end{frontmatter}  

%\date{\today} 
\maketitle   

%\newpage  
%\vspace{1.5cm}   

%%%%%%%%%%%%%%%%%%%%%%%%%%%%%%%%%%% 
\section{Introduction} 
%%%%%%%%%%%%%%%%%%%%%%%%%%%%%%%%%% 
Nonlinear one-dimensional dissipative systems at their critical points  
(like chaos threshold, bifurcation points, etc) have become one of the  
vivid areas of research since 1997, after pioneering work of  
Tsallis et al. \cite{TPZ}. One of the relevant reason for this,  
in our opinion, was that they were able to constitute one of the first  
methods of inferring the entropic index $q$ of nonextensive  
entropy \cite{ts1}   

\beq 
S_q= \frac{1-\sum_{i=1}^{W} p_i^q}{q-1}\; ,  
\eeq 
and consequently determining the proper thermostatistics which the system  
under consideration obeys, from the microscopic dynamics of the system.  
This method depends on the measure of the divergence of initially nearby  
orbits. More precisely, the effect of any uncertainty on initial conditions  
exhibits, for periodic and chaotic orbits, an exponential temporal evolution  
as  $\xi(t)=\lim_{\Delta x(0)\rightarrow 0} \frac{\Delta x(t)}{\Delta x(0)} 
\sim \exp(\lambda t)$, where $\lambda$ is the standard Lyapunov exponent,  
$\Delta x(0)$ and $\Delta x(t)$ are the uncertainties at times 0 and $t$ 
($\lambda<0$ for periodic cases, and $\lambda>0$ for chaotic ones); 
whereas at critical points (e.g., chaos threshold) this feature is 
related to a power-law type such as   

\beq 
\xi(t)= \left[1+(1-q_{mix})\lambda_{q_{mix}} t\right]^{1/(1-q_{mix})}\;\; 
\eeq 
where $\lambda_{q_{mix}}$ is a generalized version of the Lyapunov exponent  
(here, $mix$ stands for {\it mixing} and the aim of this notation will be  
transparent soon). This equation corresponds to power-law growth of the  
upper bounds of a complex time dependence of $\xi(t)$ and these upper  
bounds allow us to infer the proper value of $q_{mix}$ for the system  
under consideration. Let us call this Method I of finding the proper  
$q_{mix}$ of a given dynamical system. 

The second method (Method II) is based on the geometrical aspects of the  
critical attractor of the dynamical system at the chaos threshold.  
Scaling arguments have shown that the appropriate $q_{mix}$ is related  
to the multifractal structure of the critical dynamical attractor  
by \cite{lyra} 

\beq 
\frac{1}{1-q_{mix}} = \frac{1}{\alpha_{min}} - \frac{1}{\alpha_{max}}\;\;  
\;\;\;\;\;\;\;\;\;  (q_{mix} < 1) 
\eeq 
where $\alpha_{min}$ and $\alpha_{max}$ are the end points of the multifractal 
singularity spectrum of the critical attractor and characterize the most 
concentrated and most rarefied regions respectively on the attractor.  
This fascinating relation relates power-law sensitivity or mixing  
(left-hand side) with purely geometrical quantities (right-hand side)  
and constitutes another independent method for inferring the proper  
$q_{mix}$ of the dynamical system under study.

Finally, a third method (Method III) has been proposed for obtaining  
$q_{mix}$, dealing with the entropy increase rates \cite{latbar2,ugur1}.  
This method is the one where nonextensive entropy $S_q$ enters directly  
and consequently it provides also a justification for the use of notation  
$q$ within the two previous methods. The procedure for Method III is the  
following: We first partition the phase space into $W$ cells, then we  
choose one of these cells (randomly or not) and locate $N$ initial conditions 
all inside this cell. At any time $t$, we have $N_i(t)$ points in each cell 
(naturally $\sum^W_{i=1} N_i(t)=N$) and thus we can define a set of  
probabilities through $p_i(t)=N_i(t)/N$ ($\forall i$), which enable us to 
calculate the entropy $S_q(t)$. Using this procedure, one can define a 
generalized version of the Kolmogorov-Sinai entropy as  
$K_q\equiv \lim_{t\rightarrow\infty}\lim_{W\rightarrow\infty} 
\lim_{N\rightarrow\infty} S_q(t)/t$, and the proper $q_{mix}$ value of 
a given system is the one for which the entropy production $K_q$ is finite. 
If $q<q_{mix}$ ($q>q_{mix}$), $K_q$ diverges (vanishes).

These three independent methods (from now on referred to as 
{\it mixing-based approach}) have already been tested and verified  
with numerical calculations for a number of one-dimensional dissipative  
systems such as the logistic map \cite{TPZ}, $z$-logistic map  
family \cite{costa}, the circle map \cite{lyra}, $z$-circular map  
family \cite{circle}, asymmetric logistic map family \cite{ugur4} and the  
single site map \cite{ugur3}. Quite remarkably, it is found that all  
three methods give one and the same value for $q_{mix}$ of a given system. 
It is worth mentioning here that all these numerical findings have been 
supported very recently by an analytical work \cite{fulvio}.

We can now turn our attention to the second approach, which will be  
referred to as {\it relaxation-based approach}. This approach has been 
developed by one of us and his collaborators \cite{moura} and it is based 
on investigating numerically the critical temporal evolution of the volume 
of the phase space occupied by an ensemble of initial conditions spread 
over the {\it entire} phase space. In that sense, this approach differs 
from the mixing-based approach, where a set of initial conditions spread 
in the vicinity of the inflection point is used. In the beginning of time, 
all cells are occupied and $S_q$ is bounded by the equiprobability as 
$S_q=(W^{1-q}-1)/(1-q)$. As time  evolves, the number of occupied cells 
(say $W_{occ}$) starts contracting as 
$\left[1+(q_{rel}-1)t/\tau_q \right]^{1/(1-q_{rel})}$, here  
$q_{rel} > 1$, $rel$ stands for relaxation and $\tau_q>0$ is a 
characteristic relaxation time. In \cite{moura}, $z$-logistic map family  
has been analyzed and $q_{rel}(t\rightarrow\infty)$ values for $z$  
parameter has been obtained.

The existence of two different classes of $q$ values coming from these  
two different approaches (namely, mixing-based and relaxation-based 
approaches) is also present in the results recently obtained for fully  
developed turbulence \cite{beck,arimitsu}. Therefore, all these findings  
naturally force us to ask an intriguing question: are these classes  
connected to each other or not ? The affirmative answer to this question  
came very recently by Borges at al. \cite{ernesto}.  
Their procedure goes as follows: Beginning with $N$ points located 
inside a single cell (similar to Method III explained before), the dynamics 
of the system (for $t\to\infty$) leads it to a state with a stationary 
value of $S_{q_{mix}}$ --- this is the attractor of the system for the 
particular conditions of $W$, $N$ and the initial cell chosen. 
The path that the system evolves towards its attractor changes as the 
initial cell is changed. Some of the initial cells spread the points very fast, 
increasing the entropy to values much higher than that of the attractor 
(there is an overshooting in  the time evolution of $S_{q_{mix}}$).  
Consequently, for these specific cases, the attractor is achieved from above. 
The procedure consists in observing the rate in which the system goes to 
its attractor, for those initial cells which yield large overshoots. 
Specifically, the time evolution of 
$\Delta S_{q_{mix}} \equiv S_{q_{mix}}(t)-S_{q_{mix}}(\infty)$ is followed. 
The procedure is repeated for increasing values of $W$ (and correspondingly 
increasing values of $N$). A power-law regime appears and becomes more  
pronounced for increasing $W$. The slope (in a log-log plot) of the range  
of the $\Delta S_{q_{mix}}$ curve that follows a power-law is identified 
with  $1/(q_{rel}(W)-1)$ (with $q_{rel}>1$).  It is worth to stress that 
$q_{rel}=q_{rel}(W)$, i.e., the rate in which the  system reaches its final 
state depends on the coarse graining ($W$) adopted. The connection between 
the mixing-based and the relaxation-based approaches is established by the 
scaling law 

\beq   
\label{scaling-law}   
q_{rel}(\infty)-q_{rel}(W) \propto W^{-|q_{mix}|} \;\; .
\eeq

In this work, our aim is to provide further support to this connection  
by analyzing, for the first time, a two-dimensional dissipative system  
within this context. To accomplish this task, we will focus on the Henon  
map \cite{henon}   

\begin{eqnarray} 
\begin{array}{cc} 
x_{t+1} = \\ y_{t+1} =
\end{array}
\begin{array}{cc} 
1-ax^2_t+y_t \\ bx_t 
\end{array}  
\end{eqnarray} 
where $a$ and $b$ are map parameters. This map reduces to the standard logistic  
map when $b=0$, whereas it becomes conservative when $b=1$; in between these  
two cases, it is a two-dimensional dissipative map. Specifically, we focus on  
small values of the $b$ parameter such as $b=0.001$, $0.01$, $0.1$, etc.  
This map has been studied within the mixing-based approach using the three  
different methods and it is found that the $q_{mix}$ value for all values of  
$b$ parameter coincides with that of the logistic case, as  expected \cite{ugur}.  
Here, we first investigate the Henon map within the relaxation-based  
approach and then establish the connection between these two different  
approaches following the same lines of \cite{ernesto}.   

%%%%%%%%%%%%%%%%%%%%%%%%%%%%%%%%%%% 
\section{Numerical Results} 
%%%%%%%%%%%%%%%%%%%%%%%%%%%%%%%%%% 
As introduced in the previous Section, for the relaxation approach,  
we follow the dynamical evolution of an ensemble of initial conditions  
uniformly distributed over the phase space. In practice, a partition  
of the phase space on $W$ cells of equal size is performed and  
a set of $N$ initial copies of system is followed whose initial  
conditions are uniformly distributed over the entire phase space.  
Our results for the temporal evolution of the volume occupied by the  
ensemble are given in Fig.~1 for various $b$ values.  
As it is evident, after the power-law contraction of the volume 
occupied by the ensemble sets up, the slope [$\mu=-1/(1-q_{rel})$] is 
independent from the $b$ values and coincides with that of the logistic 
map (namely, $\mu=0.71$ hence $q_{rel}=2.41$).

Now we are at the position to establish the connection between this  
result and the one obtained in \cite{ugur}.  
Fig.~2 shows the temporal evolution of $S_{q_{mix}}$ ($q_{mix}=0.2445$) 
for different values of $W$, at the edge of chaos for $b=0.1$. 
(It would be recommended that $N \gg W$, but we adopted here $N=W$,  
due to restricted computer capabilities. Of course, we compared this case 
with $N=10\,W$, and we found that the differences are negligible.) It becomes 
evident that the limits $\lim_{W\to\infty}$ and $\lim_{t\to\infty}$ are 
non-commutable. If the former is taken first, the system will never relax to 
a stationary state (equilibrium is but a particular case), which leads to 
the conclusion that reaching equilibrium depends 
on the coarse graining adopted, and ultimately, depends on the observer! 
Equilibrium is only possible with incomplete knowledge ($W<\infty$). 
To arrive to the scaling law stated by Eq.~(\ref{scaling-law}), we must take  
first the temporal limit, and then $W\to\infty$. 
When we consider the downhill side of these curves and take the slopes of  
$\Delta S_q$ versus time (as explained earlier), then we end up with Fig.~3, 
where the verification of the scaling (4) is evident. 
On the other hand, the extrapolation value of $q_{rel}(W\to\infty)$ is 
computed as $\approx 2$, which is different from the logistic case 
($\approx 2.4$). This discrepancy may be originated, most probably, from the 
fact that $b=0.1$ value is too large for analyzing, as it is also observed 
from the deterioration seen in Fig.~1 of \cite{ugur} (Method I) for increasing 
values of the $b$ parameter. A careful investigation of the effect of 
increasing $b$ values to $q_{rel}(W\to\infty)$ would be a subject of an 
extensive computational effort. 

%%%%%%%%%%%%%%%%%%%%%%%%%%%%%%%%%%% 
\section{Conclusions}  
%%%%%%%%%%%%%%%%%%%%%%%%%%%%
After the analysis of two-dimensional dissipative Henon map at the chaos 
threshold using three distinct methods (Methods I, II and III) \cite{ugur}, 
it became evident that the scenario, which we called here 
{\it mixing-based approach}, is valid not only for one-dimensional dissipative 
map families, but also for two-dimensional cases. At this point, what 
missing in the literature were (i)~to study the {\it relaxation-based 
approach} for a two-dimensional dissipative system making use of the 
same procedure applied so far to one-dimensional cases \cite{moura}, 
and (ii)~to verify the connection (given by the scaling relation (4)) 
between the mixing-based and relaxation-based approaches for a 
two-dimensional dissipative system.
This work constitutes the first step along this line and provides 
some further elements that reinforce the idea of weak mixing.

%%%%%%%%%%%%%%%%%%%%%%%%%%%%%%%%%%   
%%%%%%%%%%%%%%%%%%%%%%%%%%%%%%%%%%% 
\section*{Acknowledgments} 
%%%%%%%%%%%%%%%%%%%%%%%%%%%%%%%%% 
This work has been supported by the Turkish Academy of Sciences, in the  
framework of the Young Scientist Award Program (UT/TUBA-GEBIP/2001-2-20).  
%%%%%%%%%%%%%%%%%%%%%%%%%%%%%%%%%%%%%%%%%%%%%%%  

%\newpage   
%%%%%%%%%%%%%%%%%%%%%%%%%%%%%% 

\newpage  

%\vspace{1.5cm}  

{\bf Figure Captions}  

\vspace{1.5cm}  

{\bf Figure 1} - Time evolution of the volume occupied by the ensemble for  
various values of $b$ parameter. Thick line corresponds to a straight line  
with $\mu=0.71$.\\   

{\bf Figure 2} - Time evolution of $S_{q_{mix}}$, with $q_{mix}=0.2445$ 
(log-log scale). 
The Henon map is at its chaos threshold for $b=0.1$, that is, 
$a_c=1.26359565...$. Points correspond to $W=700 \times 700$, 
$1000 \times 1000$, $2000 \times 2000$, $4000 \times 4000$ and 
$5000 \times 5000$. \\  

{\bf Figure 3} - Scaling law for the Henon map  
(Eq.~(\protect\ref{scaling-law})), at the chaos threshold with $b=0.1$.  
Correlation coefficient is $\approx0.99$.  To improve the correlation,  
it becomes necessary to use greater values of $W$ and $N$, that means 
better computer machinery. Dashed line indicates extrapolated value of  
$q_{rel}(W\to\infty)\approx1.987$ (that is different from the value of the  
logistic ($b=0$) map: $q_{rel}(W\to\infty)\approx2.4$). \\  

\begin{figure}
%\vspace{30mm}
\includegraphics[scale=0.6]{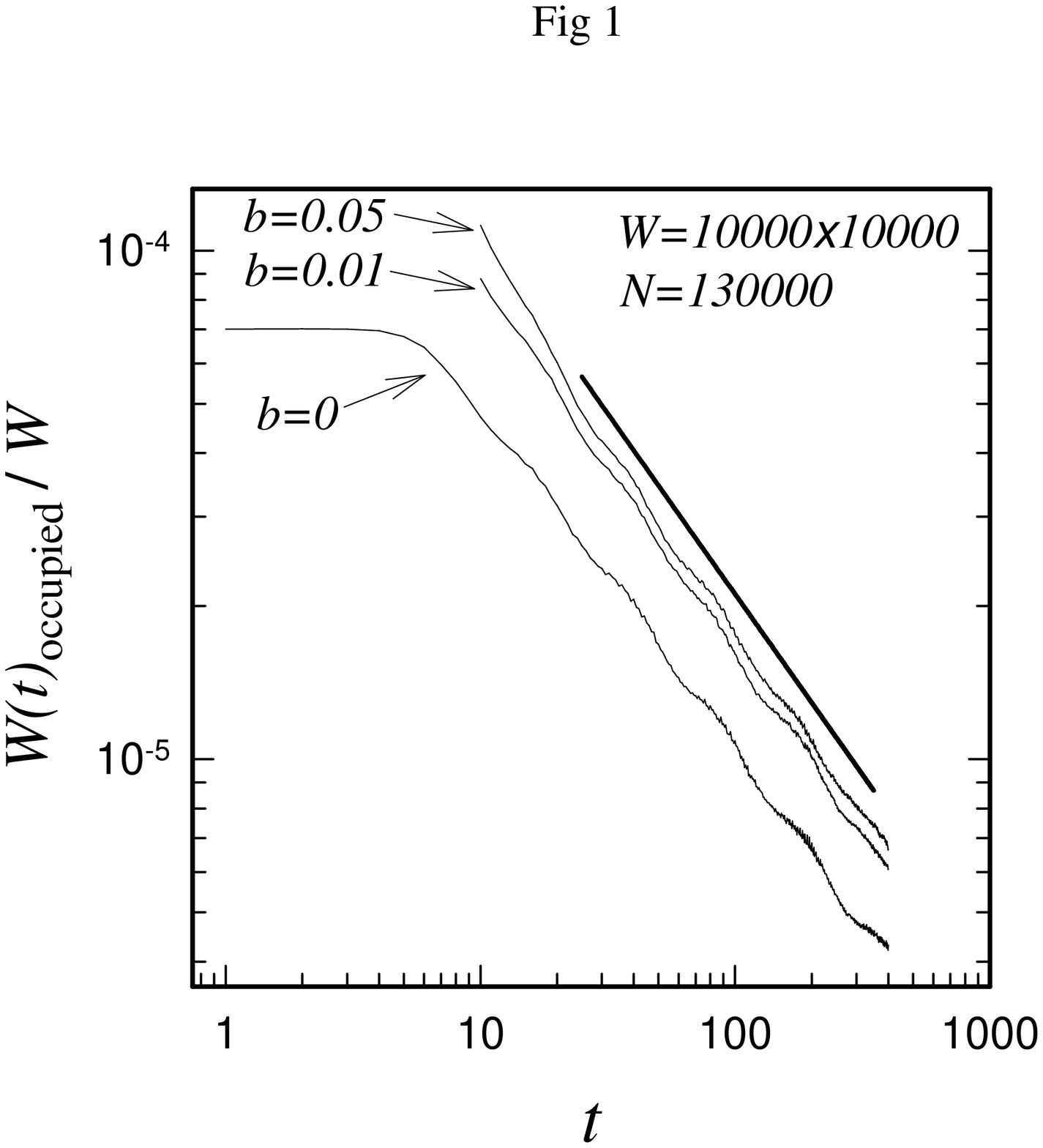}
%\caption{Caption is here...}
\end{figure}

\begin{figure}
%\vspace{30mm}
\includegraphics[scale=0.65]{tirfig2.eps}
%\caption{Caption is here...}
\end{figure}

\begin{figure}
\vspace{90mm}
\includegraphics[scale=0.7]{tirfig3.eps}
%\caption{Caption is here...}
\end{figure}

\end{document}